\documentclass[prb,showpacs,floatfix,superscriptaddress,amsmath,amssymb,twocolumn]{revtex4-1}
\usepackage{graphicx}
\usepackage{graphics}
\usepackage{dcolumn}
\usepackage{bm}
\usepackage{pstricks}
\usepackage{amssymb}
\usepackage{hyperref}
\usepackage{soul}

\definecolor{lineadecolor}{rgb}{0.35,0.5,0.6}
\definecolor{ddcol}{rgb}{0.8,0.1,0.1}
\definecolor{subsectioncolor}{rgb}{0.1,0.01,0.5}

\definecolor{celeste}{rgb}{0.8,0.87,0.99}

\def\nn{\nonumber}
\newcommand{\ba}{\begin{eqnarray}}
\newcommand{\ea}{\end{eqnarray}}
\def\be{\begin{equation}}
\def\ee{\end{equation}}

\def\spin{\vec{\mathbf{S}}}

\begin{document}

\title{Dimerized ground states in spin-$S$ frustrated systems.}

\author{ C.A.\ Lamas}
\affiliation{IFLP - CONICET. Departamento de F\'isica, Facultad de Ciencias Exactas. Universidad Nacional de La Plata,C.C.\ 67, 1900 La Plata, Argentina.}

\author{ J.M.\ Matera}
\affiliation{IFLP - CONICET. Departamento de F\'isica, Facultad de Ciencias Exactas. Universidad Nacional de La Plata,C.C.\ 67, 1900 La Plata, Argentina.}
\affiliation{Institute for Theoretical Physics, University of Ulm, Albert-Einstein-Allee 11, 89069 Ulm, Germany}

\begin{abstract}
We study a family of frustrated anti-ferromagnetic spin-$S$ systems 
with a fully dimerized ground state.
Starting from the simplest case of the frustrated zig-zag spin ladder, we generalize the family to more complex geometries like tetrahedral ladders and
 spin tubes. 
After present some numerical results about the phase diagram of these systems,
we show that the ground state is robust against the inclusion of weak disorder in the couplings as
well as several kinds of perturbations, allowing to study
some other interesting models as a perturbative expansion of the exact one. 
A discussion on how to determine the dimerization region in terms of quantum information estimators
is also presented. 
Finally, we explore the relation of these results with a the case of the a 4-leg spin
tube which recently was proposed as a model for the description of the
compound Cu$_2$Cl$_4$·D$_8$C$_4$SO$_2$, delimiting the region of the parameter space where this model
presents dimerization in its ground state.
\end{abstract}

\maketitle

\section{Introduction}

%
%
%

The Majumdar-Gosh Model (MGM)  \cite{MG} is one of the  paradigms in the physics of one dimensional spin chains.
This model represents one of the first examples of systems with gapped spectrum whose ground state (GS) is exactly known.
Although the exact result is valid only in the point $J_{2}=J_{1}/2$, it results representative of an extended phase, a dimerized phase.
A dimerized phase corresponds to a non-magnetic phase without ``classical analog'', characterized by short range, strong quantum correlations.
Systems presenting this kind of phases have gained interest in the quantum information community since, from a technological point of view, this
kind of correlations could be exploited as a resource for quantum information processing\cite{NC.00,VeCi.04,VPC.04}.
On the other hand, from a conceptual perspective, models presenting this kind of non-classical phases provide a rich playground to explore the relations between frustration and entanglement \cite{TGI.03,SuWaLi.05,
LMM.11,MGI.13}. 


%
Since its discovery, lots of efforts were devoted looking for extensions of this model to different coupling configurations\cite{SS.81,FSOF.12},
as well as more general lattices \cite{Ge.91,*BG.93,Xi.95,*BB.01,*TM.02,*Mila-exact,*Kivelson-exact}, and larger values of the local spin $S$\cite{KM.08,*Rachel.09,*MVM.12,*Penc-exact}. 
Generalizations of the MGM  to come up with more realistic models are also extremely important for the theoretical description of frustrated magnets like Cs$_2$CuCl$_4$\cite{cscucl}, KCuCl$_3$\cite{TlCuCl}, TlCuCl$_3$\cite{TlCuCl},
NH$_4$CuCl$_3$\cite{NhCuCl}, etc, which are being currently under experimental investigation.  

In this paper we present a family of anti-ferromagnetic spin $S$ models with an exact dimer product state as its ground state. 
The family includes some of the aforementioned models given a generalization to larger $S$  without the need to include quartic or more complex
many-body terms in the Hamiltonian\cite{KM.08,*Rachel.09,*MVM.12,*Penc-exact}. 

As in the case of the MG model, where the GS represents an extended phase, the manifold in the parameter space where the GS can be analytically determined is representative of an extended dimerized phase, 
that covers a large region of the parameter space.  

Unlike traditional phases, which can be characterized in the framework of the Ginzburg Landau Theory, novel phases like topological or quantum spin liquids can not be characterized in terms of local order parameters and broken symmetries.
In this way, quantum information measures have been proved to be a useful tool to characterize them\cite{ZaPa.06,*YoYi.07,*AHZ.08}, giving also information about the structure of the state and its correlations.
For this reason, in this work we employ both measures of likelihood as the global and local fidelities to the fully dimerized state, as well as pairwise and block measures of quantum entanglement.
Despite these quantities are not easily accessible from direct measures, they can be estimated from experimental parameters by measures of the structure factors\cite{*KKBBKM.09,*CPW.2011,*MaMo.13,MEKBPC.14}
or magnetic susceptibilities\cite{*WVB.05,*CSDTSC.12,KoKw.15}.

With these tools, we show that the fully dimerized state can be seen as the starting point to characterize non magnetic phases like such one observed in  KCuCl$_3$\cite{TlCuCl}.
This material has a zigzag structure of Cu-ions and corresponds to a particular limit of the model studied in the present work. This kind of quasi-one dimensional antiferromagnet structures has been intensively 
studied\cite{FSOF.12,SNT.96,SB.04,WeOi.98,Sato.07}.

Later, we use this family as a building block to construct out more complex models with a similar ground state.
In particular, we show that the ground state of a family of frustrated four-leg spin tubes is also a product
state of singlets. This dimerized phase may appears in the strong coupling regime in three leg spin tubes\cite{KaTa.97}.
Besides, we analyze the relationship between these exactly solvable models and the case of frustrated spin tubes,
which has been recently proposed as a model for the magnetic behavior of the compound Cu$_2$Cl$_4$·D$_8$C$_4$SO$_2$.
This compound seems to present frustrating anti-ferromagnetic next-nearest-neighbor exchange\cite{experimental_tube_prl,experimental_tube_prb}
and inelastic neutron scattering experiments  reveal that it presents gapped and strongly one-dimensional excitations\cite{tube_inelasting_scatering}. 
So far, there are very few theoretical studies of this frustrated model. By means of numerical analysis we show in this work that the ground state of this
model presents similar features that those found for the exactly solvable case.

The paper is organized as follows. In Section \ref{sec:exgs}, we present the family of anti-ferromagnetic frustrated ladders and show that its ground state is
a fully dimerized state. Then, a discussion about the spectrum of these systems and the magnetic behavior is presented.
In Section \ref{sec:vecinity} the properties of the ground state in the vicinity of the exact dimerizing condition are discussed 
in terms of some quantum information correlation measures.
Afterward, we show that the exactly solvable model is representative of an extended manifold.
In Section \ref{sec:tubes}, the previous  model is used as a building block for more complex systems which also have a fully dimerized ground state. 
For the particular case of frustrated spin tubes, the manifold in the phase space with partially  dimerized GS is explored by means of numerical analysis.
A numerical study of the ground state 
corresponding to the effective model of the compound Cu$_2$Cl$_4$·D$_8$C$_4$SO$_2$ and a comparison
with the exact ground state of a very similar model, belonging to the family where the ground state
can be analytically determined, are performed.
Finally, in Section \ref{sec:conclusions} the conclusions and some perspectives are presented.

\section{Exact ground state in spin-S ladders}
\label{sec:exgs}
\subsection{Exact manifold}
\label{sec:exact-lines}
We consider the following Heisenberg model on a two legs spin-$S$ ladder
\ba
\label{eq:general_Hamiltonian}
\nn
\mathbf{H}&=&\sum_{i=1}^{N} J(i)\; \spin_{2i-1}\cdot\spin_{2i}
+J'(i)\; \spin_{2i}\cdot\spin_{2(i+1)-1}\\
&+& J''(i)\; \spin_{2i-1}\cdot\spin_{2(i+1)}
+J_{2}(i)\; \spin_{2i}\cdot\spin_{2(i+1)}\\\nn
&+& J_{2}(i)\; \spin_{2i-1}\cdot\spin_{2(i+1)-1} \,,
\ea
where $\spin_{k}$ represents the local spin on the site $k$ and $\spin_{k}\equiv \spin_{k+2N}$,
with $N$ being the number of rungs. 
This ladder is represented in Figure \ref{fig:X-ladder}-A. Let us consider the case where all
the couplings are positive (all the interactions are anti-ferromagnetic). Starting from the general spin-$S$
model without translational invariance in Eq. (\ref{eq:general_Hamiltonian}) 
we can show that, imposing a simple constraint on the couplings on each square plaquette
defined by the sites $\{2i-1,2i,2(i+1)-1,2(i+1)\}$, the ground state of the system is the fully dimerized state
$ |{\psi}\rangle=\bigotimes_{i=1}^{N}|0\rangle_{i}$, with
\begin{eqnarray}
\label{eq:dimerstate}
 |0\rangle_{i}= \frac{1}{\sqrt{2 S+1}}\sum_{m=-S}^{S}(-1)^{m+S}|m,-m\rangle_{i},
\end{eqnarray}
where index $i$ labels the rung in the ladder and 
$|m,-m\rangle_{i}$ are product states such that ${\bf S}^{z}_{2 i-1}|m,-m\rangle_{i}=-{\bf S}^{z}_{2 i}|m,-m\rangle_{i}=m |m,-m\rangle_{i}$ on the rung $i$.
In order to show that $|\psi \rangle$ results an eigenstate of ${\bf H}$ with energy
$E_{0}=-J N S(S+1)$,
we rewrite the Hamiltonian in terms of local operators on each rung
\begin{eqnarray}
\vec{\bf L}_{i}&=&\vec{\bf S}_{2i}+\vec{\bf S}_{ 2i-1}\label{eq:JopDef} \\
\vec{\bf K}_{i}&=&\vec{\bf S}_{2i}-\vec{\bf S}_{2i-1} \label{eq:KopDef}\,.
\end{eqnarray}
Here, $\vec{\bf L}_i$ is the total angular momentum of the rung $i$,
and $\vec{\bf K}_i$ is a set of local observables which completes the full local Lie algebra of observables:
\begin{subequations}
\label{eq:localliealgebra}
\begin{eqnarray}
   \left[{\bf L}_{\mu},{\bf L}_{\nu}\right]&=& {\bf i} \epsilon_{\mu\nu\eta} {\bf L}_{\eta}\\
   \left[{\bf L}_{\mu},{\bf K}_{\nu}\right]&=&  {\bf i} \epsilon_{\mu\nu\eta} {\bf K}_{\eta}\\
   \left[{\bf K}_{\mu},{\bf K}_{\nu}\right]&=& {\bf i} \epsilon_{\mu\nu\eta} {\bf L}_{\eta}, 
\end{eqnarray}
\end{subequations}
where $\epsilon_{\mu\nu\eta}$ is the fully antisymmetric Levi-Civita symbol and ${\bf i}$ is the imaginary unit (${\bf i}^2=-1$).
In terms of these rung operators the Hamiltonian reads
\begin{eqnarray}
\nonumber
{\bf H}&=& \sum_{i=1}^N J(i)\left(\frac{{\bf L}^2_{i}}{2}- S(S+1)\right)+\\ 
\label{eq:Hamiltonian_LK} 
&+& \sum_{i=1}^N \frac{J'(i)+J''(i)+2J_{2}(i)}{4} \; \vec{\bf L}_i\cdot \vec{\bf L}_{i+1}+\\
\nonumber &+& \sum_{i=1}^N \frac{-J'(i)+J''(i)}{4} \left( \vec{\bf K}_i\cdot \vec{\bf L}_{i+1}
-\vec{\bf L}_i\cdot \vec{\bf K}_{i+1}\right)+\\\nonumber
&+& \sum_{i=1}^N \frac{-J'(i)-J''(i)+2J_{2}(i)}{4} \;    \vec{\bf K}_i\cdot \vec{\bf K}_{i+1}.     \label{hjk}
\end{eqnarray}
If $J'(i)+J''(i)=2 J_{2}(i)$, the last term in (\ref{eq:Hamiltonian_LK}) vanishes and
the state $|\psi\rangle$ is an eigenstate of the Hamiltonian due to
$\vec{\bf L}_i|0\rangle_i=0$. 
Noteworthy, this result is valid for any value of the local spin magnitude $S$.%

Now, provided the condition $J(i)>\frac{(S+1)}{2}(J^{'}(i-1)+J^{'}(i)+J^{''}(i-1)+J''(i))$,  $|\psi\rangle$, we show that it is the only
ground state of the system. For this purpose, we rewrite the Hamiltonian in a convenient form (see Figure \ref{fig:X-ladder}):
\begin{figure}[t!]
\begin{centering}
\includegraphics[width=0.8\columnwidth]{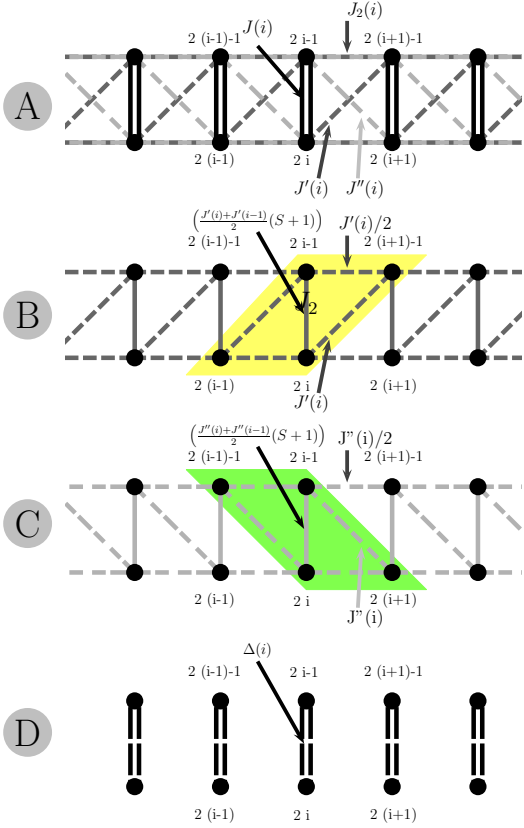}
\vspace*{-.3cm}
\caption{(Color On-line) Tetrahedral-ladder geometry and its decomposition in terms of three Hamiltonians with fully dimerized GS.}
\label{fig:X-ladder}
\end{centering}
\end{figure}

\begin{eqnarray}
\mathbf {H}&=&\mathbf {H}_{B}+\mathbf{H}_C + \mathbf{H}_D + E_D\\
\nonumber
\mathbf {H}_{B}&=&\frac{1}{4}\sum_{i=1}^N\left[ J^{'}(i-1)\left( \spin_{2i-2}+\spin_{2i-1}+\spin_{2i}\right)^2+
\right. \\ \nonumber && \left.
+J^{'}(i)\left( \spin_{2i-1}+\spin_{2i}+\spin_{2i+1}\right)^2+  
\right. \\ \nonumber && \left.
+(J^{'}(i-1)+J^{'}(i)) S\,(\spin_{2i-1}+\spin_{2i})^2-
\right. \\ \label{hAdef}  && \left.
- (J^{'}(i-1)+J^{'}(i))S(S+1)
\right] \\
\nonumber
\mathbf {H}_{C}&=&\frac{1}{4}\sum_{i=1}^N\left[ J^{''}(i-1)\left( \spin_{2i-3}+\spin_{2i-1}+\spin_{2i}\right)^2+
\right. \\ \nonumber && \left.
+J^{''}(i)\left( \spin_{2i-1}+\spin_{2i}+\spin_{2i+2}\right)^2+  
\right. \\ \nonumber && \left.
+(J^{''}(i-1)+J^{''}(i)) S\,(\spin_{2i-1}+\spin_{2i})^2-
\right. \\ && \left.
- (J^{''}(i-1)+J^{''}(i))\,S(S+1) 
\right]\label{hBdef}\\
\mathbf{H}_{D}&=&\sum_{i=1}^N\frac{\Delta(i)}{2}\left[(\spin_{2i-1}+\spin_{2i})^2\right]
\end{eqnarray}
where $E_D=-S(S+1)\sum_i J(i)$ is the energy for the  $|\psi\rangle$ state and
$$\Delta(i)=J(i)-\frac{S+1}{2}(J^{'}(i-1)+J^{'}(i)+J^{''}(i-1)+J''(i))$$
 is an effective coupling constant associated to ${\bf L}_i^2$, the total angular momentum of the pair.
$\mathbf{H}_{B,C}$ correspond, up to a constant, to the Hamiltonians of 
two zig-zag ladders (see Fig. \ref{fig:X-ladder}-B and \ref{fig:X-ladder}-C),
while $\mathbf{H}_{D}$ is the Hamiltonian of a set of uncoupled pairs (Fig. \ref{fig:X-ladder}-D). 
Now we will show that $\mathbf{H}_{B}$ and $\mathbf{H}_C$ are semi-definite positive operators, 
being $|\psi\rangle$ its ground state.

To see it, we observe that the minimum eigenvalue of a sum of operators is bounded from bellow
by the sum of the minimum eigenvalues of its terms:
$$
\min_{ \lambda\in \Lambda \left(\sum_i\mathbf{h}_i\right)}\lambda\geq\sum_i \min_{\lambda_i\in \Lambda \left(\mathbf{h}_i\right)}\lambda_i,
$$
where $\Lambda({\mathbf O})=\{\lambda_1, \lambda_2, \ldots \}$ denotes the spectrum of operator $\mathbf{O}$.
Now, we notice that for $J'(i-1),J'(i)>0$, due to the theorem of addition of angular momentum,  each term in
(\ref{hAdef}) and (\ref{hBdef}) is bounded from bellow by
{\scriptsize$$\lambda_i\geq \frac{J'(i-1)+J'(i)}{4}\min_{l_{i}=\in \mathbb{N}_0} |S-l_{i}|(|S-l_{i}|+1)-S(S+1)-S (l_{i}+1) l_{i}=0$$} and hence, 
$\mathbf{H}_{B},\mathbf{H}_C\geq 0$. Here, $l_{i}$ is the total spin in rung $i$.

On the other hand, is easy to verify that $\mathbf{H}_{B}|\psi\rangle=0$, from which it follows that $|\psi\rangle$ is an 
eigenvector of $\mathbf{H}_{B,C}$ with minimum eigenvalue. Since for each $i$, 
$J(i)>\frac{S+1}{2}(J^{'}(i-1)+J^{'}(i)+J^{''}(i-1)+J''(i))$, $\mathbf{H}_{D}$ also results positive, and hence, 
${\bf H}\geq E_{D}$. But $|\psi\rangle$ is an eigenstate of ${\bf H}$ which saturates that bound, so it is a ground state of $\mathbf{H}$. 
Due ${\bf H}_D$ is gapped, $|\psi\rangle$ is the unique state that saturates the bound.

For the $S=1/2$ case, we can improve this bound by observing that 
$\min \Lambda((\spin_{2i-1}+\spin_{2i}+\spin_{2i+1})^2)=\min \Lambda((\spin_{2i-2}+\spin_{2i-1}+\spin_{2i})^2)=3/4$,
disregarding the value of $(\spin_{2i-1}+\spin_{2i})^2$. 
This allows us to move the terms in $(\spin_{2i-1}+\spin_{2i})^2$ in $\mathbf{H}_B$ and $\mathbf{H}_C$ to $\mathbf{H}_D$, 
leading to the improved bound $J(i)>\frac{1}{2}(J^{'}(i-1)+J^{'}(i)+J^{''}(i-1)+J''(i))$.

Finally, for the translational invariant case, $J(i)=J$, $J'(i)=J'$, $J''(i)=J''$, the sufficient condition for the exact
dimerization is given by $J'+J''=2 J_2$ and 
\begin{equation}
  \label{eq:dimcond}
  J(i) > \left\{\;_{(S+1)(J'+J'')}^{(J'+J'')}\;\;_{S>1/2}^{S=1/2}   \right.
\end{equation}
The Majumdar-Ghosh point can be recovered from this result as a limit. For $S=\frac{1}{2}$, homogeneous
couplings $J\rightarrow J'=2J_2$ and $J''=0$ the GS is still a dimerized state but it is degenerate. 

\begin{figure}[t!]
\begin{centering}
\includegraphics[width=0.95\columnwidth]{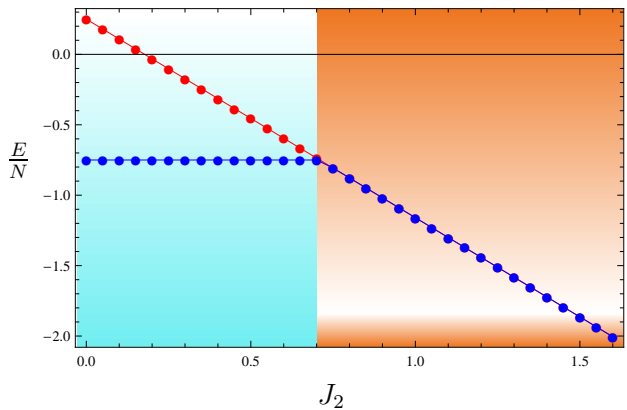}
\vspace*{-.3cm}
\caption{(Color On-line) Ground State Energy per bond corresponding to the $S=1/2$ ladder with 
$J'=J''$ (blue on-line) and the Ground State Energy corresponding to a $S=1$ chain (red on-line) as a function
of $J_{2}$ calculated with DMRG for N=60 rungs(sites). In the blue light region the ladder has a dimerized ground state
whereas in the orange region the $S=1/2$ ladder is equivalent to the $S=1$ Heisenberg chain.
}
\label{fig:en_ladder_dimer_vs_S_1}
\end{centering}
\end{figure}

\subsection{Localized triplons}

An interesting subfamily of models corresponds to the \emph{symmetric Tetrahedral ladder}, which is obtained
by setting $J'(i)=J''(i)=J_{2}(i)$\cite{Ge.91,BG.93,Xi.95,BB.01,TM.02,SSJSP.83}. 
In this subfamily,  those terms in (\ref{eq:Hamiltonian_LK}) containing operators $\vec{{\bf K}}$
vanish and the Hamiltonian depends only on the total spin of each rung $\vec{{\bf L}}_{i}$.
%
%
\begin{eqnarray}
\label{eq:effective_chain_L} 
\!\!\!\!\!\!{\bf H}=\! \sum_{i=1}^N J(i) \left(\! \frac{{\bf L}^2_{i}}{2}\!-\! S(S\!+\!1)\!\right) 
+ J'(i) \; \vec{\bf L}_i\cdot \vec{\bf L}_{i+1}.
\end{eqnarray}
\normalsize
Since $[{\bf L}_i^2,\vec{\bf L}_{j}]=0$, $[{\bf H},{\bf L}_i^2]=0$. Therefore, each eigenspace of ${\bf H}$ can be characterized by $\{l_i\}$, 
the set of total angular momentum quantum numbers (${\bf L}_i^2|\{l_i\},\ldots\rangle=l(l+1)|\{l_i\},\ldots\rangle$) associated to each rung.  Hence, on each proper subspace, the model reduces to
a spin chain with different values of the spin at each site ($l_{i}=0,1,2...2S$) with on-site quadratic terms 
$\sum_{i=1}^N J(i)\frac{\vec{{\bf L}}^2_{i}}{2}$ and exchange terms $\vec{\bf L}_i\cdot \vec{\bf L}_{i+1}$. If the total spin in a given rung is zero, there is no coupling with its neighboring rungs.  

For  $S=1/2$ the ground state of the system may correspond to $l_{i}=0$ or $l_{i}=1$, this is, the elementary excitations
of the system can be seen as localized triplons. 
The number of these triplons in the ground state is determined by the competition between the terms  $ J(i)\frac{\vec{{\bf L}}^2_{i}}{2}$ and $J_{2}(i)\vec{\bf L}_i\cdot \vec{\bf L}_{i+1}$. 
By setting the condition $(S+1) (J'(i)+J'(i-1))< J(i)$, $|\psi\rangle$ is the ground state, corresponding to $\{l_i=0\}$.
On the other hand, large values of $J_{2}$ favors larger values of $l_{i}$. In Figure \ref{fig:en_ladder_dimer_vs_S_1}
we show the energy per bond for a $S=1/2$ ladder as a function of $J_2$ with $J'=J''$ (blue circles), 
calculated by means DMRG\cite{alpscollaboration}.
Red circles corresponds to the energy of the GS in the sector $l_i=1$. It is clear that for $J_{2}>0.7$ the spin ladder behaves like a $S=1$ spin chain \cite{Xi.95,Ge.91,Honecker_jumps_xladder}.
%

This result is important for two reasons. On the one hand, it gives us a picture about how the dimerization breaks when we cross the boundaries
of the exact manifold: for large enough $J'$, the system suffers a level crossing to a higher value of the rung local spin $l_i$. On the other hand,
we can take advantage of this result to gain information about the magnetic behavior of the system.

We introduce a coupling term to an uniform external magnetic field $h$:
\begin{eqnarray}
\label{eq:effective_chain_Lb} 
\!\!\!\!\!\!{\bf H}'=\!  -h \sum_i {\bf L}_{i,z}+{\bf H}.
\end{eqnarray}

For the $S=1/2$ case the magnetic process has been studied by mean a strong coupling approach\cite{Mi.98}.
The external magnetic field induces a competition between the 
one-site term and the magnetic field term. If all  $J(i)$ are large, the GS remains being $|\psi\rangle$ up to a 
magnetic field value (lets say $h_{1}$) where there is a crossing level  and the GS becomes the state
with singlets in the even/odd rungs and fully polarized spin-1 states in the odd/even rungs. Notice that for these states the exchange interaction does not contributes. 

Then, if we increase more  the magnetic field, we obtain a second crossing level with the state containing fully polarized spin-1 states in all the rungs. In the same way, we will find successive crossing levels with fully polarized states
with  total spin $l_{odd}$ and $l_{even}$ in odd and even rungs.  
Each one of these crossing levels result in a jump in the magnetization curve followed by a magnetization
plateau. 
Hence, the magnetization process is given by successive jumps and plateaux. 
When we slightly move away from the condition $J'(i)=J''(i)$, these plateaux may remain present in
the magnetization curve, whereas the jumps are smoothed.
This frustration induced plateaux has been studied 
for $S=1/2$, $S=1$ and $S=3/2$ ladders\cite{Mi.98,Honecker_jumps_xladder,*OOS.05,*TOOS.07,Lamas-Pujol_2010,spin32-plateaux-1,*spin32-plateaux-2}
The present analysis provides a simple theoretical explanation for this behavior.

\section{Vicinity of the exact manifold}
\label{sec:vecinity}
Although we have shown that the state $|\psi\rangle$ is the GS of the model (\ref{eq:Hamiltonian_LK})
just when the condition $J'(i)+J''(i)=2J_{2}(i)$ is fulfilled, this state is representative of a region in the parameter space, where results an accurate approximation to the true ground state. 
In order to characterize the region presenting dimer order, we consider both measures of similarity between  $|\psi\rangle$ and the GS, as well as measures of entanglement.

The \emph{Uhlmann's Quantum Fidelity}\cite{NC.00} provides a measure of the similarity between two quantum states: 
$$
{\cal F}[\rho,\sigma]={\rm Tr}\sqrt{{\sqrt{\sigma}}\rho{\sqrt{\sigma}}}={\cal F}[\sigma,\rho]
$$ 
where $\rho$ and $\sigma$ are two quantum states of the same system. If $\sigma$ corresponds to a pure state $|\alpha\rangle$, this quantity reduces to $
{\cal F}[\rho,|\alpha\rangle]=\sqrt{\langle \alpha|\rho|\alpha\rangle}
$. 

We start considering the fidelity between $|\psi\rangle$ and the true GS in a region close enough to the exact manifold such that a first order perturbative treatment would be feasible.
Starting from the Hamiltonian (\ref{eq:Hamiltonian_LK}), and using the algebraic properties of ${\bf L}_{i,\mu}$ and ${\bf K}_{i,\mu}$ the canonical first order perturbation theory leads to 
\begin{equation}
  \label{eq:perturb1}
|{\rm GS} \rangle\approx  (1-\frac{3}{64}\,N \gamma^2)^{1/2}|\psi\rangle + \frac{1}{4}\sum_{i} \sqrt{\frac{3}{4}}\gamma_i  | i,i+1\rangle  \,,
\end{equation}
where $|i,j\rangle=\frac{3/4}{\sqrt{3}  S(S+1)}  \vec{\bf K}_{i}\cdot \vec{\bf K}_{j}|\psi\rangle$,
$\gamma_i=\frac{S(S+1)}{3/4}\frac{J'(i)+J''(i)-2J_2(i)}{(J(i)+J(i+1))/2}$ and 
$\gamma^2=\frac{1}{N}\sum_{i}\gamma_i^2$. Notice the explicit $SU(2)$ invariance of the approximation, 
as well as its translational invariance for the homogeneous case.  In this way, the global fidelity is given 
by $(1-\frac{3}{64}\,N\, \gamma^2)^{1/2}$, which is valid for $\frac{3}{64}\,N \gamma^2 \ll 1$. 
From now on, we are going to restrict to this last case and hence, $\gamma_i=\pm\gamma$. 
In this case, the previous result seems suggest that for large systems the
dimerization is constrained just over the exact manifold.
However, to look for high values of the fidelity in a large system is a very demanding condition.
Due to its definition, for product states ${\cal F}[|\alpha\rangle^{\otimes N},|\alpha'\rangle^{\otimes N}]=({\cal F}[|\alpha\rangle,|\alpha'\rangle])^N$
and hence, yet for very similar states, the fidelity vanishes in the large $N$ limit. 
On the other hand, from (\ref{eq:perturb1}) we can estimate the fidelity for the state of a single rung ($\rho_{12}$) against the singlet state
${\cal F}_{0}[\rho_{12}]={\cal F}[\rho_{12},{|{\rm singlet}\rangle}]\approx (1-\frac{3}{64}\gamma^2)^{1/2}$. 
When this approximation is valid, we can see that ${\cal F}[|{\rm GS}\rangle,|\psi\rangle] \approx {\cal F}_{0}[\rho_{12}]^N$, which is just what we
expect if the global state behaves like a product of the local states of the rungs, which is an important
feature of the dimerized phase. 
A similar result can be obtained by the method of variational cluster mean field + RPA discussed in \cite{Matera-Lamas_2014}. This treatment predicts that the GS is well approximate by $|\psi\rangle$ plus small Gaussian correlations
for $|\gamma|<1$. 

Now, we will see that for the $S=1/2$ case,  the value of ${\cal F}_{0}[\rho_{12}]$ determines
most of the relevant features of the dimerized phase. 
To see this, we observe that due to the $SU(2)$ symmetry, for the $S=1/2$ case the local state $\rho_{ij}$ of a subsystem composite by the (single spin) sites ($i,j$) is
completely determined by ${\cal F}_0[\rho_{ij}]$:
\begin{figure}[t!]
  \centering
   \includegraphics[clip,width=0.99\columnwidth]{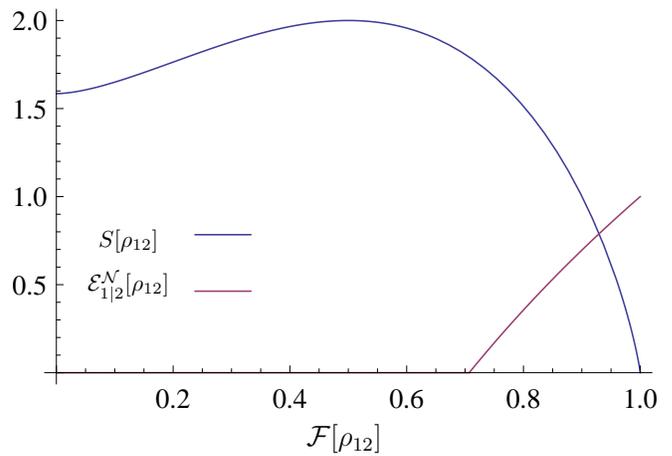}
  \caption{(Color On-line) Relation between entanglement and fidelity to the singlet state for $SU(2)$ invariant states of pairs of $S=1/2$ systems. For ${\cal F}_{0}[\rho]<1/\sqrt{2}$ the internal degrees of freedom are not entangled. 
}
  \label{fig:fidelityent}
\end{figure}
\begin{equation}
  \label{eq:rhoS1hpairstate}
\rho_{ij}= \frac{1-{\cal F}_0^2[\rho_{ij}]}{3}{\bf 1}_4+
\frac{4{\cal F}_0^2[\rho_{ij}]-1}{3}|0\rangle\langle 0| \,.  
\end{equation}
Since $\rho_{ij}$ is the state of a subsystem associated to pure global state, a measure of the correlations between this subsystem and the rest of the system
is given by its entanglement entropy\cite{NC.00}. For the state (\ref{eq:rhoS1hpairstate}) it is reduced to
$$
S(\rho_{ij})=h\left({\cal F}_0^2[\rho_{ij}]\right)+3
h\left(\frac{1-{\cal F}_0^2[\rho_{ij}]}{3}\right)
$$
where $h(x)=-x \log_2(x)$. For the dimerized phase, where the global state is well approximated by the fully dimerized state, which is a product state,
$S(\rho_{12})$ should remain  small, while $S(\rho_{1 j})\approx 2$ for $j\neq 2$.
In order to give a more accurate idea about the limit value of ${\cal F}_0[\rho_{12}]$ for which it makes sense to talk about dimerization, it 
is better to analyze the internal degree of entanglement of the rung. 
A measure of the degree of pairwise entanglement for mixed states is provided by the logarithmic negativity\cite{VW.02,Pl.06}
$$
{\cal E}^{\cal N}_{\cal{A}|\cal{B}}[\rho]=\log_2 |\rho^{\rm t_{\cal{A}}}|_1
$$
 where $|A|_1={\rm tr}\sqrt{A^{\dagger}.A}$ is the trace norm (or the Schatten $- 1$ norm) of the matrix $A$, and  $\rho^{t_{\cal A}}$ represents the partial transposition of $\rho$ with respect to the subsystem ${\cal A}$, i.e. the linear 
 map defined by $(|\alpha\rangle_{\cal A}|\beta\rangle_{\cal B}\langle\alpha'|_{\cal A}\langle\beta'|_{\cal B})^{t_{\cal A}}=|\alpha'\rangle_{\cal A}|\beta\rangle_{\cal B}\langle\alpha|_{\cal A}\langle\beta'|_{\cal B}$. 
 This quantity is saturated by ${\cal E}^{\cal N}[|{\rm singlet}\rangle]=\log_2(1+2 S)$ and vanishes for every separable state. 
For the state (\ref{eq:rhoS1hpairstate}) ${\cal E}^{\cal N}$ is given by
$$
{\cal E}^{\cal N}_{i|j}[\rho_{ij}]=\log_2(1+\max(0,2{\cal F}^2_0[\rho_{ij}] -1)) \,.
$$

In Figure \ref{fig:fidelityent},  the behavior of $S[\rho_{12}]$ and ${\cal E}^{\cal N}_{1|2}[\rho_{12}]$ as a function of ${\cal F}_0[\rho_{12}]$ is depicted.
Notice that for ${\cal F}_0[\rho_{12}]\leq \frac{1}{\sqrt{2}}$, the reduced state  is separable  and hence,  the correspondent global state is not dimerized anymore.

\begin{figure}[t!]
  \centering
   \includegraphics[clip,width=1.\columnwidth]{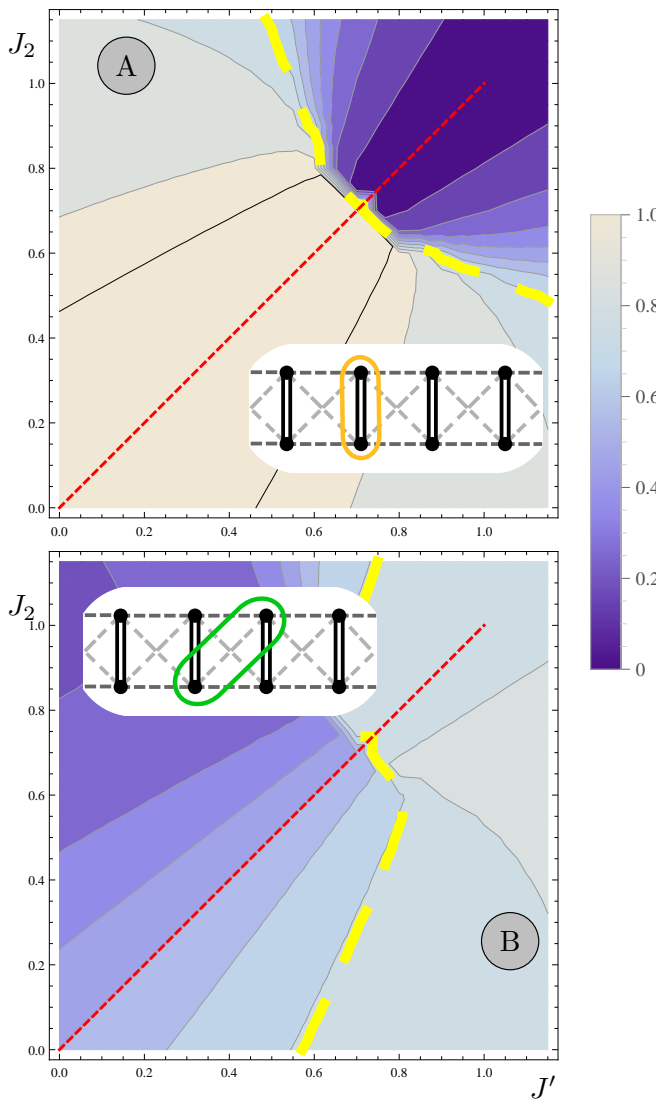}
  \caption{(Color On-line) Landscape of the fidelity between the singlet state and the state of the internal pair $\rho_{12}$ (A)
and state of the nearest neighbor external pair $\rho_{23}$  (B) associated to the 
ground state for the spin  $1/2$ ladder with $J'=J''$ for $N=8$ rungs. The straight dashed line (red on-line) indicates the exact dimerized line $J_{2}=J'$. 
In the panel (A), the dark continuous line indicates the boundary of the region with fidelity $>0.95$.
In both panels, the thick dashed curve (yellow on-line) corresponds to the value $1/\sqrt{2}$, limiting the region where the correspondent local states are entangled.
Insets indicate the pair used to calculate the fidelity.
}
  \label{fig:fidelity}
\end{figure}

In Figure \ref{fig:fidelity} a landscape of the fidelity between the singlet state and the state of the rung ${\cal F}_0[\rho_{12}]$ (A) and 
a nearest neighbour external pair  ${\cal F}_0[\rho_{23}]$ (B),
obtained by numerical evaluation is shown, for the case of the symmetric ladder $(J''=J')$. The presented results were
evaluated by means of the Lanczos method\cite{alpscollaboration}.
The straight dashed line (red online) indicates the intersection with the exactly dimerized manifold where the fully dimerized state
is the ground state of the system. For the strong pair (A), the fidelity is symmetric regarding the exchange between $J'$ and $J_2$. The graphic reveals a wide dimerized region ($F_0[\rho_{12}]>0.95$) around 
the exact line $J'=J_2$, implying that the pertubative analysis is accurate over this region.  
As we cross the critical value $J'\approx 0.6 J$, $F_0[\rho_{12}]$ is suddenly reduced, due to the GS is now
orthogonal to the dimerized one.
For pairs of spins coupled by $J'$ (B), we observe that near the exact line $F_0[\rho_{23}]\approx 1/2$, 
which is consistent with a fully mixed state. For larger $J'$ these pairs become more entangled,
at the expense of the entanglement in the rung ($1-2$).  
On the other hand, increasing $J_2$ the spins on the pair $1-3$ tends to align, which reduces its fidelity to the singlet below $1/2$.

Further improvements can be obtained by mean of a cluster mean field + RPA expansion as we have shown
in a previous work\cite{Matera-Lamas_2014}. This opens a way to solve some effective models associated
to certain quasi-one dimensional materials like KCuCl$_3$.
%
%
%
%
%
%

\section{Frustrated four-leg spin tubes.}
\label{sec:tubes}
\begin{figure}[t!]
\begin{centering}
\includegraphics[width=0.8\columnwidth]{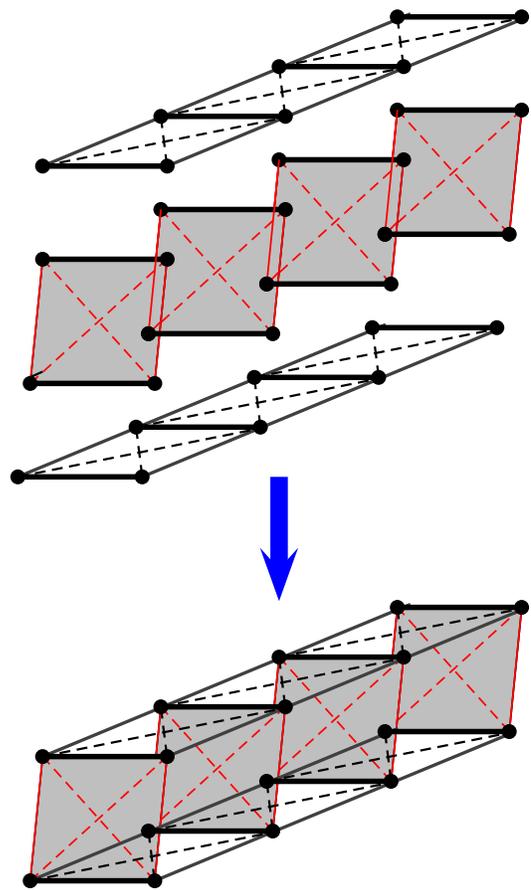}
\vspace*{-.3cm}
\caption{(Color On-line) Building a spin tube family with a fully dimerized ground state.}
\label{fig:ladder-tube}
\end{centering}
\end{figure}

The family of ladders presented above can be used as a ``building block'' to obtain more complex models
with a product state as its ground state.
It is straightforward to show that there is a family of Hamiltonians corresponding to
four-leg spin tubes that present also a dimerized ground state. 
These Hamiltonians can be written as a sum of two ladders whose Hamiltonians are given by (\ref{eq:general_Hamiltonian})
and $N$ square plaquettes as represented in Figure \ref{fig:ladder-tube}.
\ba
\mathbf{H}_{tube}=\mathbf{H}_{ladder 1}+\mathbf{H}_{ladder 2}
+\sum_{j=1}^{N}\mathbf{H}_{square, j}
\ea
Where $N$ is the number of rungs in the ladders. $\mathbf{H}_{ladder 1}$ and $\mathbf{H}_{ladder 2}$ are Hamiltonians
corresponding to the upper and lower ladders in Fig. \ref{fig:ladder-tube} and  $\mathbf{H}_{square, j}$
is the Hamiltonian of the $j$-th square plaquette in the figure. These square plaquettes have also a
dimerized ground state corresponding to dimers in the strongest bonds, 
as can be easily seen since the square is a special case of ladders discussed previously with 2 rungs.
 Let us concentrate in the homogeneous case $J(i)=J$, $J'(i)=J'$ and $J''(i)=J''$, $\forall_i$,
 keeping in mind that all the conclusions can be easily generalized for the inhomogeneous case. 
 Different combinations of couplings $J'$, $J''$ and $J_{2}$ preserving the constraint $J'+J''=2J_{2}$
 gives different geometries of four-leg spin tubes with a fully dimerized ground state that can be
 written as $|\psi_{tube}\rangle = |\psi\rangle_{ladder1} \times |\psi\rangle_{ladder2} $. 
 Then, following the same steps as in the previous section is easy to show that the state $
 |\psi_{tube}\rangle$ is an eigenstate of the system and there is a range of couplings where
 this state is the ground state of the system.
 Besides, as in the ladder case,  we can expect this exact GS to be representative of a finite region around the
 exact manifold of the parameter space.

\begin{figure}[t!]
\begin{centering}
\includegraphics[width=1.0\columnwidth]{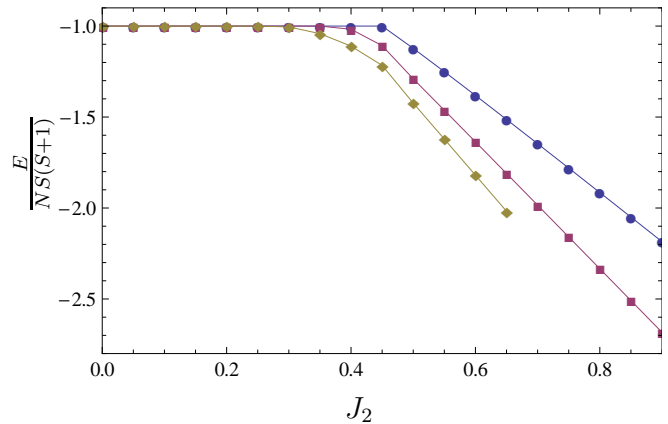}
\vspace*{-.3cm}
\caption{(Color On-line) Ground state energy per bond in units of $S(S+1)$ as a function of $J_{2}$ calculated 
with DMRG for a four-leg spin tube with 160 spins, $J''=0$ and $J'=2 J_{2}$. 
Blue circles, red squares and yellow rhombi correspond to $S=1/2$, $S=1$ and $S=3/2$ respectively.}
\label{fig:energy_exact_tube}
\end{centering}
\end{figure}

We will consider now the particular cases of tubes with $J'=J''$ and $J''=0$. For these two cases,
the fully dimerized state $|\psi_{tube}\rangle$ is the ground state of the system if the following
condition is satisfied
\begin{equation}
  \label{eq:sufficient_cond_tube}
J'<\left\{
  \begin{array}{c}
\xi J \hspace{.5cm} S=\frac{1}{2}\\
\xi \frac{J}{(S+1)}\hspace{.5cm} S\geq 1
 \end{array}\right.
\end{equation}
where $\xi=1/3$ for $J'=J''$ and $\xi=2/3$ for $J''=0$.
 In Figure \ref{fig:energy_exact_tube} the energy per rung in units of $S(S+1)$ 
as a function of the coupling $J_{2}$, corresponding to the case $J''=0$ and $J'=2J_2$ is depected
for different values of the spin.
The range of values where the ground state is the product singlet state ({\it i.e.} $E/N_{rungs}=-1$)
is bigger than that found analytically.
The reason is that Eq. (\ref{eq:sufficient_cond_tube}) represents just a sufficient condition, being the real 
range larger in general.

As we have seen for the ladders, the symmetrical case $J'=J''$ is special. 
The excited states correspond to localized triplons and the eigenvalues of the Hamiltonian can be labeled by
 the set of values of the total momentum in the rungs $\{l_i\}$. 
 This situation gives a rich magnetization profile, containing a sequence of jumps
 and plateaux. 
 Such scenario is due by the competition between the terms in the Hamiltonian
 that favors states with minimal total momentum in each rung,
 spin-exchange terms and the magnetic field contribution.
The crossover between the ground state $|\psi_{tube}\rangle$ and the state with alternation of one singlet and one triplet
in each rung determines the jump in the magnetization curve between the plateaux at $m=0$ 
and $m=\frac{1}{4S}$.

In this case, system frustration promotes the magnetization plateaux and
the jumps in the magnetization curve but also makes the ground state simpler (in each 
magnetization sector the ground state is a direct product of rung-states).

Even while the condition $J'+J''=2J_{2}$ is hold, for $J''\neq J'$ this is not true any more.
%
Changing the value of $J''/J'$ the magnetization plateaux reduce its widths and
jumps between plateaux transform into a smooth piece of the magnetization curve. 

\begin{figure}[t!]
\begin{centering}
  \includegraphics[width=0.99\columnwidth]{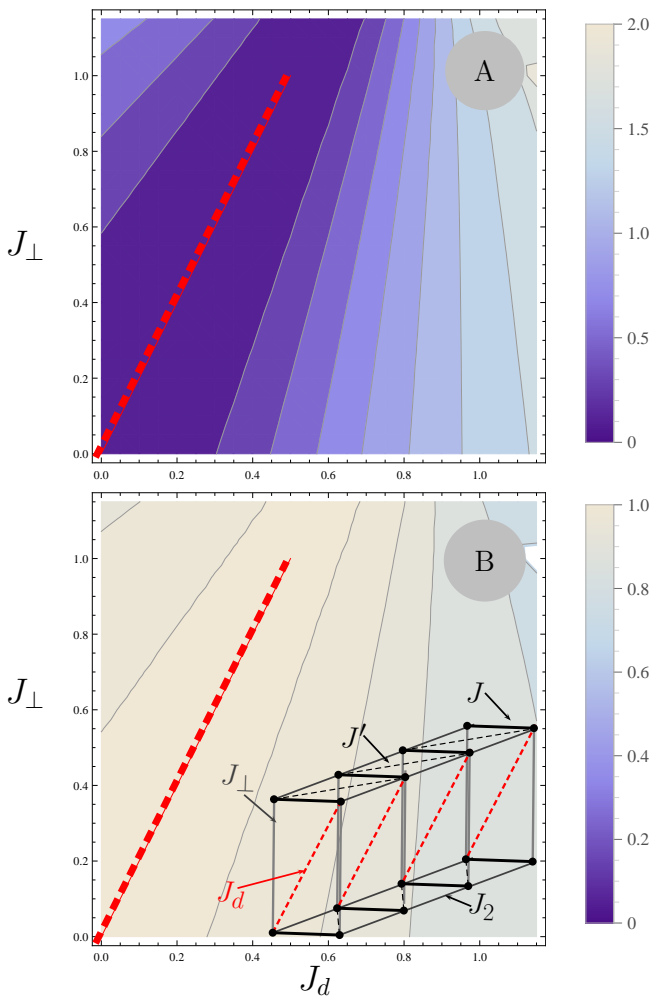}
\vspace*{-.3cm}
\caption{(Color On-line) Dimerization in the case of weakly coupled ladders as a function of inter-ladder couplings calculated by using the Lanczos method 
 with $L=4$, $J_2=0.5 J'$ and  $J'=0.95 J$ (see the inset) . 
(A) Entanglement entropy of a strong coupled pair with the rest of the tube. 
(B) Fidelity of the local state with the singlet state. The region in which the fidelity is over $0.99$ matches 
with such that the entanglement entropy is lower that $0.2$. The dashed thick line (red on-line) corresponds to the exact dimerization condition $J_{\perp}=2 J_d$
}
\label{fig:tube_inter_w}
\end{centering}
\end{figure}

The existence of the exact result for tubes also leads us to ask in which conditions, two ladders which present dimer order 
in its ground state, conserve it when they become weakly coupled, assembling a tube. From a perturbative argument, for small enough inter-ladder couplings,
we expect that the global state stays dimerized. However, due to the exact result Eq. (\ref{eq:sufficient_cond_tube}), despite it becomes more frustrated, for larger couplings the system approach to another exact dimerized configuration. 
As a result, the region presenting dimer order would be enlarged. As an example, we will consider the case in which the interaction between ladders are given between correspondent spins on each ladder and on one of the diagonals of each
plaquette (see the inset of Figure \ref{fig:tube_inter_w}), calling $J_{\perp}$ and $J_{d}$ the respective coupling constants. 
This case is interesting since recently, a similar topology was proposed (but for $J_d=0$) as the appropriate model describing the compound Cu$_2$Cl$_4$D$_8$C$_4$SO$_2$ \cite{experimental_tube_prl,experimental_tube_prb}.
Despite this case does not satisfy the exact dimerization condition for tubes, it is interesting to find out if it could support dimer order.  
In Figure \ref{fig:tube_inter_w} the behavior of the fidelity between the state of a rung and the singlet state,
as well the entanglement entropy of this subsystem with the rest of the tube is depicted, for weakly coupled zig-zag ladders, as a function of the inter-chain couplings.
Notice that the dimerization over the lateral ladders is not broken for quite large values of 
$J_{\perp}$ and $J_{d}$ near the exact dimerization condition. It would suggest that we can expect the presence of dimer order
in the model proposed for Cu$_2$Cl$_4$D$_8$C$_4$SO$_2$.  In the next sections we extend this result for a more realistic case, when the ladders does not satisfies the exact dimerizing condition. 
For it, we take advantage of the exact result for the four legs frustrated tube discussed above starting from a highly frustrated system but with a separable GS.

 \subsection{Effective Hamiltonian of the four-leg spin tube material Cu$_2$Cl$_4$·D$_8$C$_4$SO$_2$}
\label{sec:material}
\begin{figure}[t!]
\begin{centering}
\includegraphics[width=0.8\columnwidth]{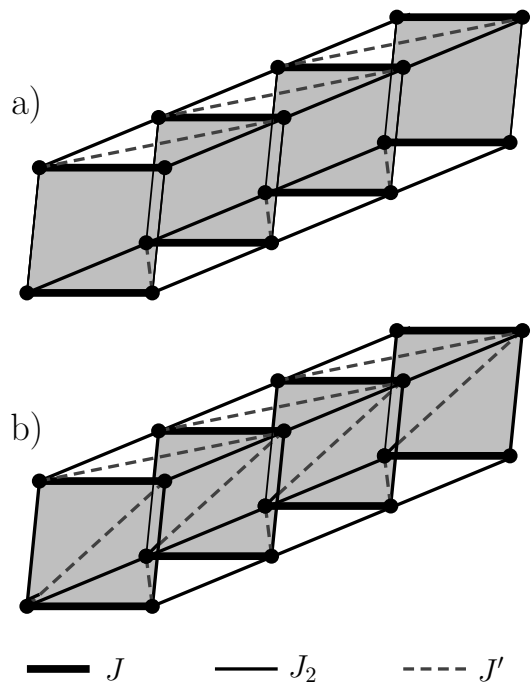}
\vspace*{-.3cm}
\caption{(Color On-line) Different spin tubes geometries.}
\label{fig:tubes_veriety}
\end{centering}
\end{figure}

In a Recent experiment, inelastic neutron scattering has been used to investigate
the magnetic excitations in the quantum spin-liquid system 
Cu$_2$Cl$_4$·D$_8$C$_4$SO$_2$\cite{experimental_tube_prl,experimental_tube_prb}.
In that work, it  was suggested that the appropriate Heisenberg Hamiltonian 
is a $S=1/2$ four-leg spin-tube with no bond alternation as the showed in Figure \ref{fig:tubes_veriety}-a).
There are a scarce number of theoretical results on this kind of prototypical models on the spin
tubes\cite{Oshikawa-tube,Arlego-Brenig-tube}.
We study numerically the frustrated four leg spin tube model proposed to describe the
compound Cu$_2$Cl$_4$·D$_8$C$_4$SO$_2$.

The proposed model is closely related to the family of spin
tubes presented in the previous sections.
Consider the member of the family of spin tubes with a fully dimerized ground state
schematized in Figure \ref{fig:tubes_veriety}-b).
This model can be obtained from the Hamiltonian proposed for the material adding an extra diagonal coupling in
each square.
This modified Hamiltonian belongs to the family of spin tubes presenting a dimer product ground state.
If the ground state properties of these two models are similar, the effective model
for Cu$_2$Cl$_4$·D$_8$C$_4$SO$_2$ may be studied starting
from the exactly known ground state, taking the diagonal couplings in the squares as a perturbation. 
Although this perturbative study is out of the scope of the present paper we can see that the dimerized state is robust in 
the exact model. This robustness suggest to make
an expansion around the dimer ground and incorporating triplon excitations.
In the rest of this section, we are going to analyze numerically the proposed model for the material Cu$_2$Cl$_4$·D$_8$C$_4$SO$_2$, looking for fingerprints on the
properties predicted for the exactly solvable case, leaving the analytical study of the corrections for a future work.

\subsection{The model for the material and the exactly solvable model}

\begin{figure*}
  \centering
\hspace*{-.5cm}\includegraphics[width=\textwidth,angle=-90,clip]{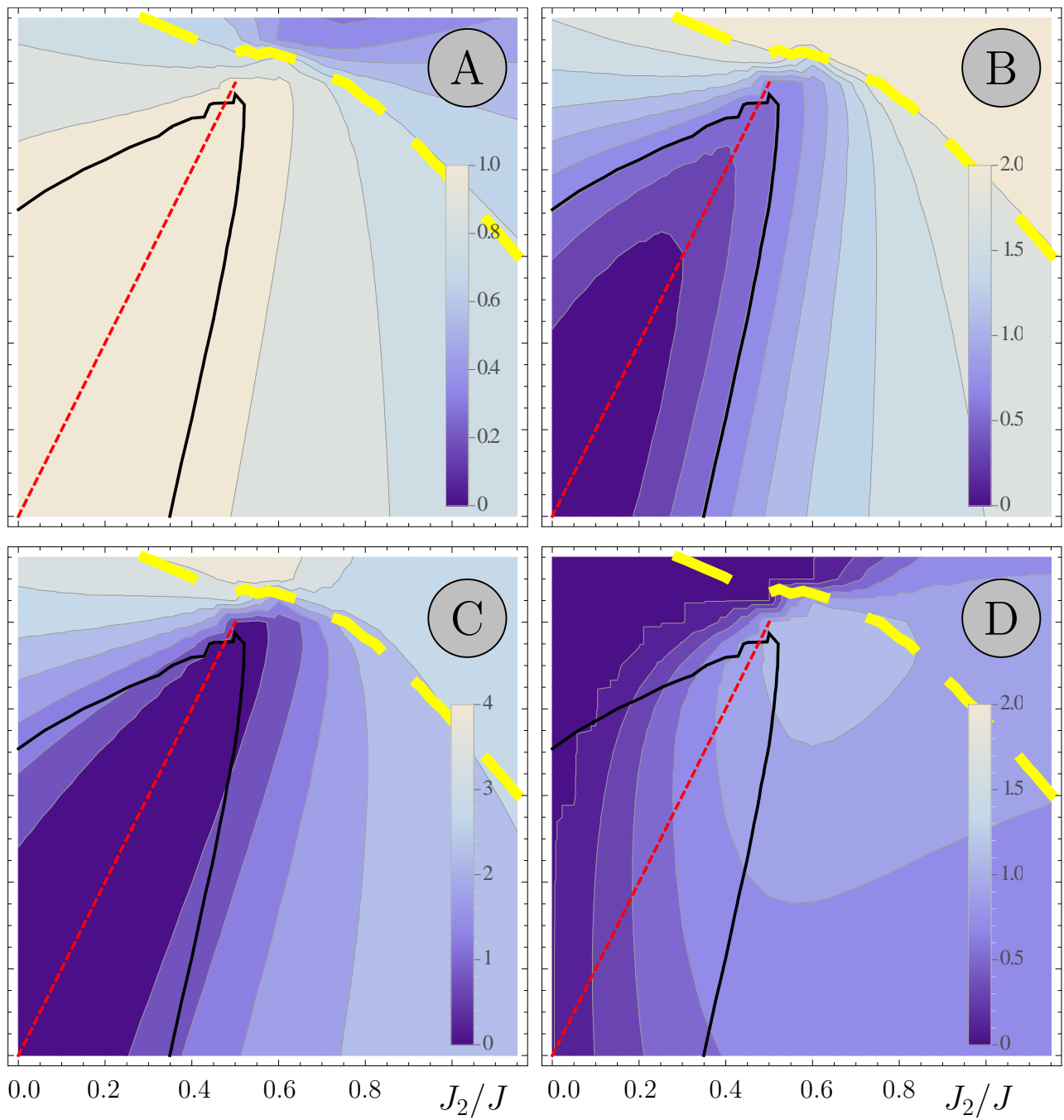}\\
\caption{(Color On-line) Local structure of the ground state for the tube, as a function of couplings $J_2$ and $J'$,
evaluated through exact diagonalization (Lanczos method) for the case $L=4$ plaquetes ($16$ sites) and $S=1/2$. 
Panel A: Fidelity of the local state associated to a single rung respect to the singlet state.
Panel B: entanglement entropy of a rung  with the rest of the tube.
Panel C: Entanglement of a plaquette; 
Panel D: internal entanglement, measured by its logarithmic negativity between two pairs in the same plaquette.
The dashed (red on-line) straight line represents the condition $J'=2J_2$; the dark continuous curve bounds the region where ${\cal F}_{0}[\rho_{12}]\geq 0.95$ while the dashed light curve (yellow on-line) is 
the bounds the region where the pair $\rho_{12}$ is entangled (${\cal F}_{0}[\rho_{12}]> 1/\sqrt{2}$).  
}
\label{fig:tubematent}
\end{figure*}
Above we have analyzed the behavior of the ground state in tubes near the dimerization condition.
Now, we are going to discus which are the common features between these results and the behavior of a
more realistic family of tubes. In particular, we consider the model proposed by  Garlea et al. in \cite{experimental_tube_prl,experimental_tube_prb} for the Cu$_2$Cl$_4$·D$_8$C$_4$SO$_2$ compound. 
In order to see what features are shared between our exactly solvable model and the model proposed in
\cite{experimental_tube_prl,experimental_tube_prb},
we explore numerically the ground state of Hamiltonians on the family
$J>J_2,J'$, corresponding to the tube in the figure  \ref{fig:tubes_veriety}-a, 
looking for common features.

In Figure \ref{fig:tubematent} a landscape of several entanglement observables, corresponding to the ground 
state of a tube with couplings as in Figure \ref{fig:tubes_veriety}-a are depicted. 
In the top panels, the fidelity between the local state of a single rung and the singlet state (Panel A)
and the entanglement entropy of a rung (Panel B) are shown.
Due to the $SU(2)$ symmetry,  for local states of a single rung the entanglement entropy is a function of the fidelity to the singlet state.
As we could expect from a composite mean field treatment\cite{Matera-Lamas_2014,BRCM.15},
the local state of a single rung can be accurately approximated by a pure singlet state in a relatively wide 
region around the condition $J'=2 J_2$, even if $J_2$ is moderately large.

On the bottom panels we can appreciate  the entanglement of a \emph{plaquette} 
composed by two parallel rungs with the rest of the tube
(Panel C) and the internal entanglement between two rungs in the same plaquette, measure by its \emph{logarithmic negativity}\cite{Pl.06} (Panel D). 
For large $J_2$ and $J'$, we observe that both the entanglement between the state of a square and the rest of the tube becomes larger, which is compatible with a symmetry broken phase. In fact, the limit $J\rightarrow 0$ with $J'$ and $J_2$ 
fixed corresponds to a non frustrated tube, for which a Ne\'el-like phase is expected. 
We can also notice that the region where the squares are not entangled is larger than the dimerized region. In particular, near $J_2\approx 0.5 J$ and $J'\approx 0.9 J$, we can observe a region where the entanglement of
the square with the rest of the tube is small, but the internal entanglement between rungs is near to $1$. This seems to indicate that such region corresponds to a resonant plaquette order, where the GS is well approximated
by  $|GS\rangle\approx  |\alpha\rangle_{square}^{N}$,
being $|\alpha\rangle_{square}$ a four spin singlet state (${\bf J}_{square}^2 |\alpha\rangle_{square}=0$), which is not a product state of the rung states $|\alpha\rangle_{square} \neq |{\rm single}\rangle|{\rm singlet}\rangle$. 

In summary, we saw that although the exact dimerization condition for this model is only possible in the trivial limit $J'=J_2=0$, the phase diagram looks quite similar to that associated to the family containing the solvable model. This can be understood  by considering the removal of the extra bond on the square hamiltonian as a perturbation over the Hamiltonian of the solvable tube. However, near the crossover point, the perturbation theory is no longer valid, which can  give place to new features in the phase diagram.

\section{Discussion and perspectives}
\label{sec:conclusions}

In the present paper, a general $SU(2)$ invariant quantum spin-S Heisenberg ladder was investigated. 
A sufficient condition for the existence of a fully dimerized exact ground state was shown for a wide
subfamily of such systems.  Besides, by means of a combination of numerical and analytical techniques,
the existence of this phase for a general  value of the local spin was proved,
showing that the region in the parameter space corresponding to the dimerized phase is reduced as
the  magnitude of the local spin grows. 
 
For the case of symmetrical frustration the excitations were also exactly determined and a 
discussion about the magnetization
process\cite{Mi.98,Honecker_jumps_xladder,*OOS.05,*TOOS.07,Lamas-Pujol_2010,spin32-plateaux-1,*spin32-plateaux-2} was presented. 

The ground state properties around the exact manifold in the parameter space were explored by means of numerical analysis, showing that these remain close to the exact case over a finite region. 
Due to the large stability of this phase against external perturbations, a quantum simulator that could reproduce this kind of couplings would be able to
 prepare a large number of fully entangled pairs in a robust way. In the last years several proposals
 for the experimental simulations of spin systems in ion trap experiments \cite{BSkRP.11,BSkRP.12,BR.12,ISC.13} suggest that this kind of setup could be readily in a near future.

Besides, we have shown how the family of Hamiltonians with a fully dimerized ground state can be extended from the family of ladders to 
more complex models.  As an example a family of frustrated 4-leg spin tubes with a dimerized ground
state was built.
Afterward, common features found in the ground state of the solvable family and
those obtained for more realistic models were analyzed. In particular, a comparison to the  model proposed
for the Cu$_2$Cl$_4$·D$_8$C$_4$SO$_2$ compound \cite{experimental_tube_prl,experimental_tube_prb} was discussed.  

We hope this study can be taken as a starting point for a more systematic study of the 
ground state properties and excitations around the lines on the parameter space
where the ground state was exactly determined.

As a perspective, the study of  hole dynamics on a background of dimers in one-dimensional
 systems \cite{BG.93}. In higher dimensions there is an important amount of results
 on dimerized ground states\cite{SSPB.81,Bo.92,MRKU.00,TJS.02,Ku.02} that can be used as
 starting point to study hole dynamics.
In 2D has been recently proved that the quantum statistics of holes in a dimer background can be changed
without affect the energy dispersion\cite{Lamas-Ralko_2012} and the density of holes has an impact on
the magnetization plateaux \cite{Lamas-Pujol_2010}. As we start from these families of models where the ground state
is a dimer covering, introducing holes in the system may result in a very interesting phase diagram.


\section*{Acknowledgments}
We acknowledge useful discussions with M. Plenio,  D.C. Cabra., P. Pujol and R. Rossignolli.
C. A. Lamas is supported by CONICET (PIP 1691) and ANPCyT (PICT 2013-0009)
and J. M. Matera is supported by CONICET and the Institute for Theoretical Physics, University of Ulm.

\vspace{1cm}
\bibliographystyle{aipnum4-1}
\bibliography{Referencias_exact_gs}


\end{document}